\documentclass[letterpaper]{article}

\usepackage[T1]{fontenc}

\usepackage{geometry}
\usepackage{setspace}

\usepackage[
    style = chem-acs,
    doi = true,         
    url = false,        
    eprint = false,     
    articletitle = true 
]{biblatex}

\usepackage{graphicx}
\usepackage{float}
\newfloat{scheme}{htbp}{los}
\floatname{scheme}{Scheme}
\floatname{chart}{Chart}
\newfloat{graph}{htbp}{loh}

\usepackage{chemformula} 
\usepackage[version = 4]{mhchem} 

\usepackage{hyperref}
\usepackage{authblk}
\author[1]{Caio C. Quaglio-Gomes}
\affil[1]{Departamento de Física, Universidade Federal de São Carlos, 13565-905 São Carlos, SP, Brazil}

\author[2,3]{Stefan Marinković}
\affil[2]{National High Magnetic Field Laboratory, Los Alamos National Laboratory, 87545 Los Alamos, NM, USA.}
\affil[3]{Experimental Physics of Nanostructured Materials, Q-MAT, CESAM, Université de Liège, B-4000 Sart Tilman, Belgium}

\author[1]{Elijah A. Abbey}

\author[4]{Davi A. D. Chaves}
\affil[4]{Centro Brasileiro de Pesquisas Físicas, 22290-180 Rio de Janeiro, RJ, Brazil}

\author[5]{Anna Palau}
\affil[5]{Institut de Ciència de Materials de Barcelona, ICMAB-CSIC, Campus de la UAB, 08193 Bellaterra, Catalonia, Spain}

\author[3]{Alejandro V. Silhanek}

\author[6]{Pedro Schio}
\affil[6]{Brazilian Nanotechnology National Laboratory, Brazilian Center for Research in Energy and Materials, 13083-100 Campinas, SP, Brazil}

\author[1,*]{Maycon Motta}

\title{Probing Electromigration of Oxygen Vacancies in YBa$_2$Cu$_3$O$_{7-\delta}$ Devices by Multimodal X-ray Techniques}

\date{*Email: m.motta@df.ufscar.br}

\begin{document}

\maketitle

\begin{abstract}
 Control of oxygen vacancies by electrical currents in complex oxides such as \ce{YBa2Cu3O_{7-$\delta$}} (YBCO) has attracted considerable interest due to the relative simplicity of its implementation and its potential for both fundamental studies and the tuning of superconducting device properties. However, the structural evolution and depth-dependent effects associated with current-based techniques remain largely unexplored, particularly with respect to the connection between optical signatures and the spatial distribution of oxygen vacancies. Here, we combine nanoprobe X-ray Diffraction (NanoXRD), Cu K-edge X-ray Absorption Near-Edge Structure (XANES), X-ray Photoelectron Spectroscopy (XPS), electrical transport, and optical measurements to reveal modifications induced in YBCO microbridges by pulsed electromigration. We observe a $c$-axis expansion correlated with spectroscopic features of oxygen depletion in the Cu–O chains, and we confirm that oxygen redistribution, crystallographic changes, and copper coordination evolve consistently across techniques. Notably, the spatial profile of unit-cell expansion closely follows the optical contrast observed after electromigration, demonstrating that the different signatures capture the same underlying oxygen reordering. We further show that optical microscopy cannot reliably capture bipolar electromigration involving strong resistance modifications, as surface deoxygenation appears largely irreversible. Taken together, our findings provide a significant step toward a microscopic understanding of current-assisted oxygen migration in YBCO and establish a framework for effectively exploiting vacancy control in high-temperature superconducting devices.
\end{abstract}

\section*{Keywords}

Oxygen vacancy, \ce{YBa2Cu3O_{7-$\delta$}}, Pulsed electromigration, Lattice distortion, X-ray characterization


\twocolumn
\section{Introduction}

The perovskite cuprate \ce{YBa2Cu3O_{7-$\delta$}} (YBCO) is one of the most studied superconducting oxides, being the first one discovered with a critical temperature $T_c$ above the boiling point of nitrogen~\cite{Chu1997}. Having reached the 40-year mark since the discovery of high-$T_c$ superconductivity, YBCO as well as other cuprates and oxide superconductors have attracted interest in applications as diverse as superconducting tapes~\cite{molodyk_prospects_2023,stangl_ultra-high_2021} and integrated quantum computing systems~\cite{confalone_cuprate_2025}, due to their many correlated properties. Although the mechanism of hole-pair binding has not been completely elucidated, the role of structure and composition in the functionalities of this material family is beyond questioning \cite{Gui2021}. 

YBCO can be found in two structural forms~\cite{Jorgensen1990}: a deoxygenated, tetragonal form that behaves as an antiferromagnetic insulator, and the well-oxygenated orthorhombic unit cell, which hosts superconducting, metallic behavior~\cite{cava_structural_1990}. Being a non-stoichiometric oxide capable of hosting a continuous range of oxygen compositions, YBCO serves as a model system for exploiting the interplay of structure, composition, and electronic phase to achieve controlled metal-to-insulator (MIT)~\cite{Veal1991, janod_resistive_2015} and superconductor-to-insulator (SIT)~\cite{Semba2001} transitions. These transitions can be tuned in situ via temperature~\cite{obolenskii_localization_2006,solovjov_effect_2019}, applied electric fields and currents~\cite{Ahn2003, roadmap2026}, or more recently, through direct laser writing~\cite{Biancardi2026}. The ability to repeatedly characterize and modify samples in situ represents a powerful shortcut over conventional techniques and challenges the combinatorial approach~\cite{Jin2013b,wu_perspective_2015}. Beyond fundamental studies, the tunability of YBCO opens opportunities for emerging computing technologies, including analog computing for neural networks and memristive devices~\cite{li_review_2018,waser_nanoionics-based_2007, Rouco2025}.

In-situ control of oxygen vacancies in YBCO can thus enable new applications of high-temperature superconducting technology. The underlying principles of vacancy diffusion have been successfully described using conventional diffusion models~\cite{DeOrio2010}, leaving the door open to further implementations that rely on electrical, thermal, and even strain-induced diffusion. Electrostatic diffusion of oxygen vacancies by applied fields in YBCO is the most widely explored concept for such doping modulation~\cite{Coll2019}. Field-gated devices have been demonstrated with a metallic~\cite{Palau2018}, ferroelectric~\cite{Begon-Lours2017} and ionic-liquid gates~\cite{Perez-Munoz2017b}, among others~\cite{lorenz_2016_2016}. Recently, Alcalà \textit{et al.}~\cite{alcala_tuning_2024} reported advances in the structural characterization of devices whose oxygen profile is modulated in this manner. Their findings indicate that the YBCO under electric-field gate is percolated by YBa$_2$Cu$_4$O$_8$ phases, which correlates with reduced superconducting and conducting performance. 

An alternative approach centered on colossal current densities, in which selective electromigration (EM) of oxygen is applied to induce doping control, has attracted increasing interest due to its feasibility as a route for oxygen-content manipulation without complex fabrication steps~\cite{marinkovic_direct_2020,Collienne2022,trabaldo_mapping_2022}. Such experiments have already been demonstrated to be able to map the phase diagram of YBCO~\cite{trabaldo_mapping_2022}, and to create reversible switches~\cite{marinkovic_electromigration-driven_2024}. While the macroscopic effects of EM on the oxygen composition of YBCO films have been widely reported, it remains to be clarified how these modifications manifest at the microstructural level, such as through unit cell distortions and electronic structure reconfiguration, and whether these modifications proceed via a filamentary mechanism as reported for other memristive systems~\cite{Gunkel2025}.

In this work, we employ bulk-sensitive nanoprobe X-ray Diffraction (NanoXRD) and X-ray Absorption Near-Edge Structure (XANES) to investigate the impact of selective oxygen electromigration in YBCO on both the crystallographic structure and the copper atomic environment, which act as proxies for oxygen content. By correlating these advanced probes with electrical transport and optical microscopy, we establish direct evidence of microstructural and electronic modifications induced by oxygen migration under different experimental conditions. Our results reveal a wave-like propagation of oxygen vacancies, with deoxygenation signatures consistently identified across all techniques, particularly in the spatial profile of unit-cell expansion that matches the optical contrast observed after moderate electromigration. Complementary surface-sensitive X-ray Photoelectron Spectroscopy (XPS) measurements further confirm these modifications, supporting a unified picture of oxygen redistribution. These findings provide a new framework for understanding current-driven oxygen dynamics and open avenues for further exploration of electromigration in complex oxides.

\section{Experimental Section}

\subsubsection{Sample Fabrication}

Three $c$-axis oriented YBCO thin films were investigated in this work. Samples S1 and S3 are 100-nm-thick YBCO films grown by Pulsed Laser Deposition (PLD) on \ce{SrTiO3} substrates (see Supporting Information SI1 for details on film deposition and characterization).  Microbridges were patterned via photolithography, followed by ion-beam etching. The central bridges of S1 and S3 have lateral dimensions of 2 $\mu$m~$\times$~4 $\mu$m, as presented in Figure~\ref{fig:1}(a). To ensure robust electrical contacts, a bilayer of 5 nm Ti and 80 nm Au was deposited onto the pads by electron-beam evaporation. Sample S2 consists of a 100-nm-thick YBCO film grown by PLD on a \ce{LaAlO3} substrate. This sample was previously used to create a finite-element model of the oxygen-vacancy counter-flow, as described in Ref.~\cite{Collienne2022}. The central bridge in S2 has dimensions of 1 $\mu$m $\times$ 5 $\mu$m.

\subsubsection{Electromigration Procedure}

Pulsed electromigration experiments were performed on all samples following the procedures described in Refs.~\cite{Collienne2022,marinkovic_oxygen_2023,marinkovic_effect_2023}, using a four-point configuration for resistance monitoring. For samples S1 and S3, a sequence of 1~s current pulses with linearly increasing amplitude (up to the mA range) was applied. During each pulse, the constriction resistance was recorded as \(R_{\mathrm{max}}\), while between pulses, a small probing current (100 $\mu$A) was used to measure the resistance in both polarities, defining \(R\). The procedure was continued until a persistent change in \(R\) of the central constriction was achieved, thereby denoting an EM run. Reflection optical images were captured at the start and end of the procedure. For sample S2, multiple electromigration runs were performed using 10 s current pulses, with polarity alternated between runs, while the resistance was monitored using the same four-point configuration.


\subsubsection{Nano X-ray Diffraction}

NanoXRD experiments were performed at the CARNAÚBA (Coherent X-ray Nanoprobe Beamline) beamline at the Brazilian Sirius synchrotron radiation source, LNLS/CNPEM~\cite{Tolentino2023}. Specifically, the TARUMÃ (Tender to Hard X-ray for Submicro Analysis) station, with its broad energy range of 2.05–15~keV and a focused beam size of 200~$\times$~200~nm$^2$, was employed in association with a scanning stage, allowing spatially resolved investigation along the devices. For the analysis of the YBCO (005) diffraction peaks, a 10-keV beam energy was selected. Measurements were performed in a fixed-grazing-incidence geometry, aligned to satisfy the Bragg's law for the targeted diffraction peak, with a PIMEGA 540D area detector covering a range of scattering angles corresponding to $\Delta q = 0.4$~$\text{\AA}^{-1}$ around the Bragg condition (sample stage tilted 76$^\circ$ relative to the incident beam, and detector positioned at 30$^\circ$). The measured peaks were fitted with Lorentzian functions to accurately extract their positions (see Supporting Information SI2).

\subsubsection{X-ray Absorption Near Edge Structure}

Cu~K-edge XANES spectra were collected along the devices at the CARNA\'UBA beamline to probe chemical modifications induced by electromigration. Spectra were collected in fluorescence mode at selected points along the devices. At each point, a scan in the 8.960–9.020 keV range with a 0.5 eV energy step and an acquisition time of 0.5 s per energy point was performed using the same 200~$\times$~200~nm$^2$ beam spot and normal-incidence geometry. The polarization of X-rays was oriented within the $ab$-plane of the YBCO thin films. Due to the relatively long exposure times associated with these scans, we first performed beam-induced damage tests at the Cu K-edge to determine safe measurement conditions (Supporting Information SI3). These tests ensured that the XANES spectra could be acquired without significant beam-induced oxygen loss or other artifacts.

\subsubsection{X-Ray Photoelectron Spectroscopy}

XPS measurements were performed at the IPÊ (Inelastic Scattering and Photoelectron Spectroscopy) beamline using a 15° incidence angle and a beam spot of 12~$\times$~6~$\mu$m$^2$. Excitation energies of 930 eV and 1330 eV were used for oxygen and copper, respectively, chosen to yield kinetic energies of approximately 400 eV for electrons emitted from the main core-level peaks. For oxygen, binding energies in the 525–538 eV range were measured to probe the O~1\textit{s} core level, while for copper, spectra were collected in the 922–970 eV range to capture both the Cu~2\textit{p}$_{1/2}$ and Cu~2\textit{p}$_{3/2}$ core levels.

\begin{figure*}[!h]
    \centering
    \includegraphics[width=1\linewidth]{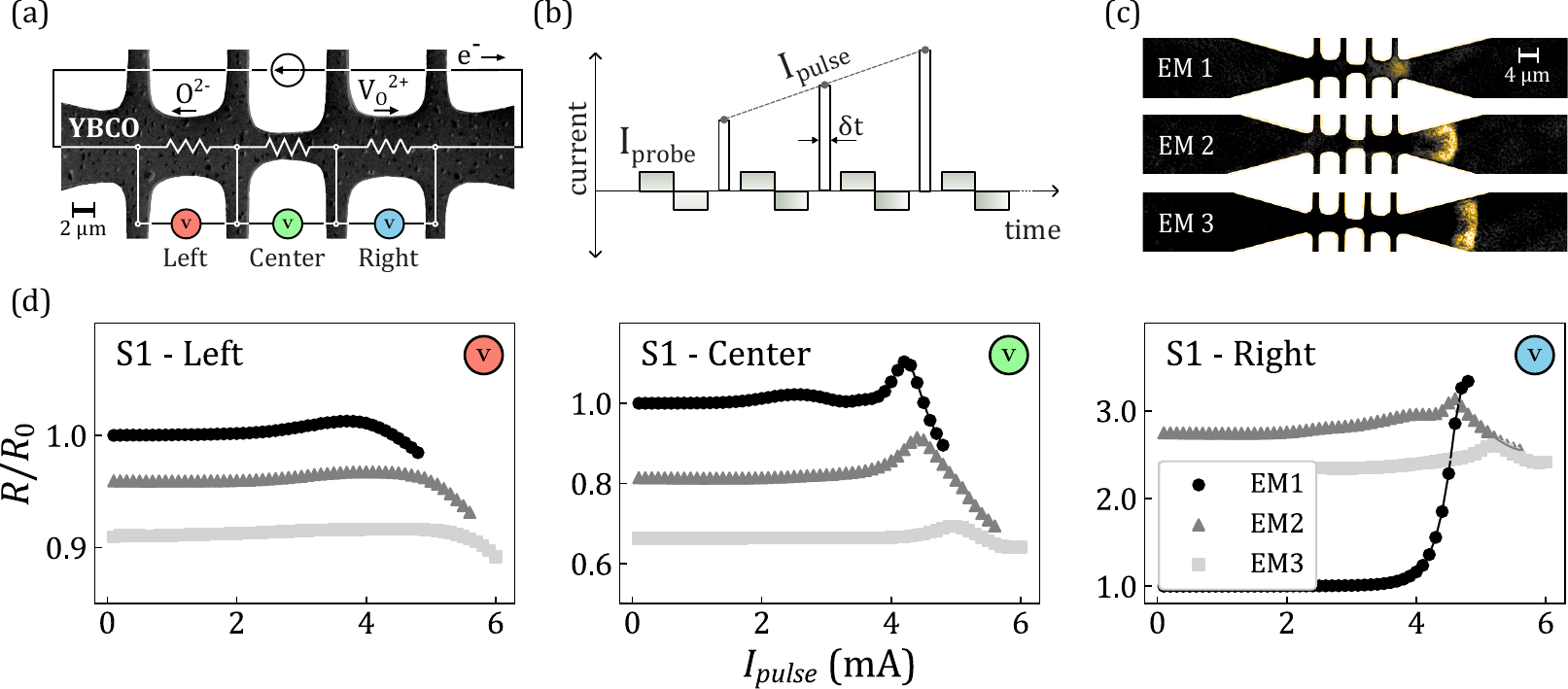}
   \caption{
(a) Schematic of the triple-constriction device with the corresponding electrical circuit layout. Arrows labeled $\mathrm{O^{2-}}$ and $\mathrm{V_O^{2+}}$ indicate the expected drift directions of oxygen anions and oxygen vacancies, respectively, under the applied current polarity. (b) Schematic of the electromigration pulse-probe current protocol (current versus time).
(c) Differential optical images of sample S1 at successive stages of electromigration, after EM1, EM2, and EM3. The enhanced contrast (yellowish regions) highlights the progressive rightward propagation of the oxygen-vacancy front.
(d) Pulsed electromigration resistance traces normalized by initial values $R_0$ for each one of the three bridges of sample S1, obtained under ambient conditions at room temperature for three consecutive EM runs.
}
    \label{fig:1}
\end{figure*}

\section{Results and Discussion}

All investigated samples present the geometric features illustrated by Figure~\ref{fig:1}(a). The sequence of three constrictions, with the innermost one being narrower than the outer ones, localizes the onset of electromigration at the central constriction due to its geometry-induced elevated current density. Concurrent with this, the outer constrictions facilitate the assessment of antisymmetric behavior resulting from oxygen-vacancy counterflow. This particular sample design has been employed in prior studies that established the methodology for current-driven selective oxygen migration in YBCO~\cite{marinkovic_direct_2020,Collienne2022,marinkovic_oxygen_2023} and related perovskite materials~\cite{marinkovic_electric_2022}.


YBCO sample S1 was subjected to a moderate process, following the pulsing protocol represented in Figure~\ref{fig:1}(b). In this device, three consecutive EM runs were performed at room temperature and atmospheric pressure with a pulse duration of 1~s. Reflection-microscopy images with micrometer-scale spatial resolution were acquired after each EM run to monitor the evolution of the optical contrast. The false-colored differential images shown in Figure~\ref{fig:1}(c) reveal the emergence of an optical-brightening front that progressively extends beyond the constriction regions from the second EM run (EM2) onward. This behavior is consistent with the expected optical response of deoxygenated YBCO~\cite{kircher1991} and provides insight into how changes in optical reflectivity correlate with oxygen movement along the bridges during electromigration.

Electrical resistance measurements provide a scenario consistent with this interpretation. Figure~\ref{fig:1}(d) shows the normalized resistance as a function of pulse-current amplitude for the left, central, and right constrictions. During the first EM run (EM1), the left and right constrictions display opposite trends characteristic of oxygen electromigration in YBCO: the left constriction decreases in resistance after EM onset ($\sim$4~mA), which is commonly associated with oxygen enrichment, while the resistance in the right constriction increases due to deoxygenation. This spatial asymmetry arises from the selected current polarity, which drives both hole transport and oxygen-vacancy drift, leading to oxygen accumulation on the left and vacancy buildup on the right bridge. In subsequent EM runs (EM2 and EM3), continued pulsing drives the device into a regime in which a deoxygenation front propagates across the structure, causing all three constrictions to exhibit decreasing resistance, consistent with electromigration behavior reported in the literature~\cite{Collienne2022}.


\begin{figure}[!b]
    \centering
    \includegraphics[width=\linewidth]{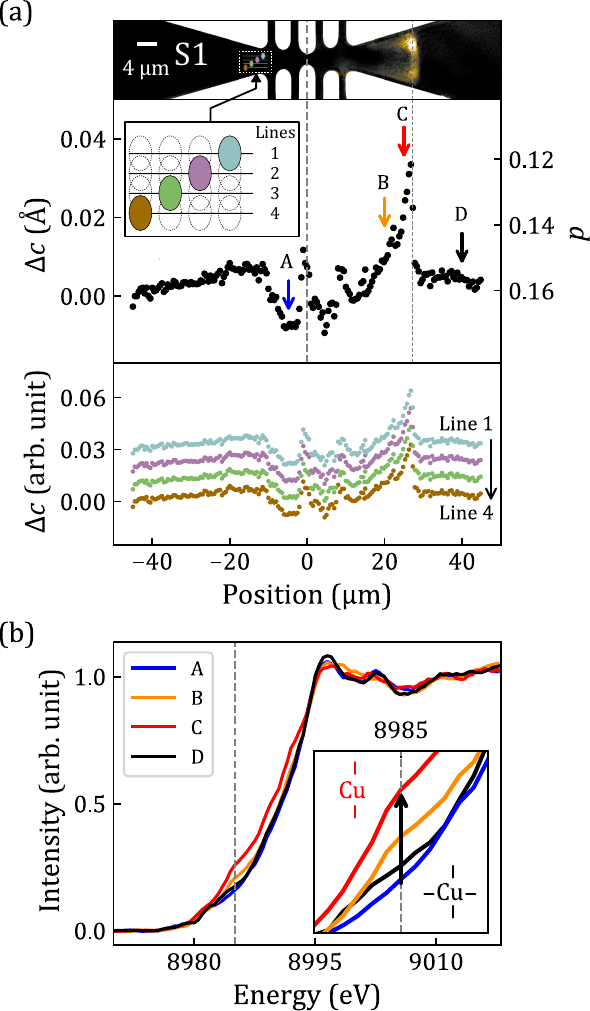}
   \caption{(a) Profile of the $c$-axis lattice parameter variation measured by NanoXRD for sample S1. The upper panel shows a false-colored optical image of the device, aligned with the NanoXRD sampling positions. The plotted curve represents the average of four parallel line scans performed along the device length, as illustrated in the inset; the four individual scans, which have been arbitrarily shifted along the y-axis for clarity, are displayed in the lower panel. The right y-axis indicates the corresponding hole doping $p$. (b) XANES spectra acquired from the regions marked by the colored arrows in (a). The inset highlights the pre-edge region, used for the qualitative assessment of oxygen content.}
    \label{fig:2}
\end{figure}

XRD is widely used to investigate how oxygen content affects the crystallographic structure of YBCO \cite{Jorgensen1990, Benzi2004}. In this work, we exploit the well-established correlation between oxygen stoichiometry, hole doping, and the $c$-axis lattice parameter to evaluate oxygen distribution. The removal of O(1) atoms from Cu–O chains alters the vertical position of the apical oxygen O(4), leading to an elongation of the $c$-axis length. Figure~\ref{fig:2}(a) shows the lattice parameter variation ($\Delta c$) of sample S1 after electromigration, determined from shifts of the YBCO (005) peak measured in the NanoXRD experiments. Each point represents the average of four measurements acquired across the microbridge width. The four parallel line scans along the device length (bottom frame of Figure~\ref{fig:2}(a)) show no significant differences, apart from a geometric effect. Because the sample is tilted relative to the incident beam, the bridge appears geometrically distorted along one direction in the beamline reference frame. The measurement points, represented schematically as colored ellipses in the inset of the figure, have centers separated by $\sim$60~nm along the projected (compressed) direction across the bridge width, which is smaller than the beam spot. As a result, the spots partially overlap, covering approximately 88\% of the width of the central bridge in this distorted projection. Under these conditions, no evidence of filamentary behavior is observed, and the lattice-parameter modulation appears wave-like along the device.

A clear spatial correlation emerges between the resistance and optical trends discussed above and the unit-cell modifications observed along the device. The left constriction exhibits a shortened $c$-axis lattice parameter, whereas the right side of the device shows a pronounced $c$-axis expansion (points A and C, respectively, in Figure~\ref{fig:2}(a)). In the central constriction, both expanded and contracted regions are observed, predominantly near the vertical voltage lines where current crowding is enhanced, consistent with the nonmonotonic behavior in the resistance. Notably, the maximum structural modification coincides with the spatial extent of the high-reflectivity region, evidencing a direct relationship between the two deoxygenation signatures. The right y-axis of Figure~\ref{fig:2}(a) shows the corresponding hole doping $p$, defined as the hole concentration per copper atom in the CuO$_2$ planes, estimated using the empirical relation proposed in Refs.~\cite{Liang2006, arpaia2018}. Overall, these results demonstrate that synchrotron-based XRD enables spatially resolved tracking of electromigration effects at the nanometer scale, going beyond what can be inferred from resistance measurements and directly linking oxygen-vacancy accumulation to local changes in crystal structure.

\begin{figure}[!b]
    \centering    \includegraphics[width=\linewidth]{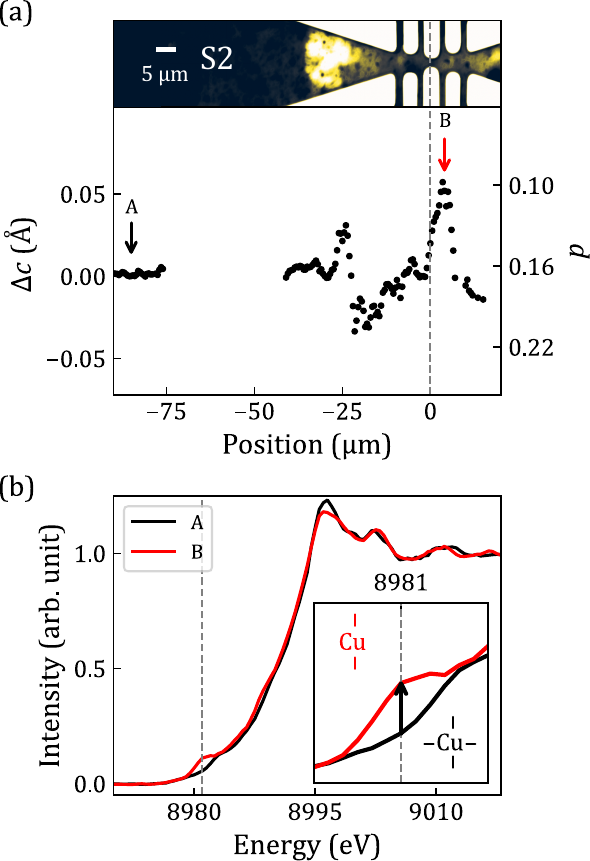}
    \caption{(a) Optical image and NanoXRD profile of sample S2. The pronounced modulation of the $c$-axis lattice parameter evidences oxygen redistribution, with the right y-axis indicating the corresponding hole doping $p$. Local hole doping spans approximately from the underdoped to the overdoped regimes. (b) Cu K-edge XANES spectra acquired at the indicated positions highlight the correlation between structural modifications and copper coordination, manifested by an increased pre-edge intensity.}
    \label{fig:3}
\end{figure}

Cu K-edge XANES was used to investigate modifications in the electronic structure induced by electromigration in the YBCO device, complementing the structural information obtained from NanoXRD. XANES provides direct insight into the local copper coordination and oxidation state, with detailed interpretations under irradiation-induced disorder given in Ref.~\cite{nicholls2022}. Four representative locations were selected based on the $c$-axis profile in Figure~\ref{fig:2}(a), covering both EM-affected regions and pristine reference areas: (A) a $c$-axis minimum at the left bridge, (B) a point in the rising region, (C) near the maximum lattice expansion, and (D) a pristine reference outside the optical contrast. Increases in the pre-edge shoulder below 8994 eV indicate local oxygen loss, consistent with previous reports \cite{Perez-Munoz2017b, nicholls2022}, as deoxygenation shifts copper coordination from square planar to linear, enabling transitions to nonbonding 4p$_y$ orbitals and enhancing the pre-edge absorption \cite{Tolentino1989}. The corresponding XANES spectra (Figure~\ref{fig:2}(b)) show a consistent trend: region A has low pre-edge intensity, indicating relative oxygen enrichment; regions B and C exhibit increasing pre-edge features correlating with lattice expansion and local deoxygenation; and region D displays low pre-edge intensity, characteristic of optimally-oxygenated YBCO. Together, the data reveal a clear correspondence between structural and electronic modifications, demonstrating the effects of electromigration-induced oxygen-vacancy counterflow.


Extending our analysis to a more extreme case, we resort to sample S2. This legacy device, which has undergone several EM runs, including both positive and negative polarity, and previously studied in Ref.~\cite{Collienne2022}, exhibits a markedly different regime compared to S1, primarily characterized by more severe EM effects. As shown in Figure~\ref{fig:3}(a), the $c$-axis lattice parameter profile displays a pronounced wave-like modulation with large-amplitude peaks and valleys, reflecting successive migration and counter-migration of oxygen vacancies. Importantly, the corresponding hole concentration $p$ (right-hand axis) reaches nearly the full extent of the superconducting dome \cite{Keimer2015}, locally accessing both the underdoped and overdoped regimes, highlighting that electromigration can be exploited to probe the cuprate phase diagram. Given the narrower central constriction of this sample, only a single NanoXRD linescan was acquired along the bridge.

In agreement with the structural variations, we also detect an enhancement of the pre-edge feature in the Cu K-edge XANES spectra in the region of the most expanded unit cell (point B in Figure~\ref{fig:3}(a)), once again indicating the copper coordination evolving from square planar toward a linear configuration. Notably, in this case, regions of maximum \textit{c}-axis expansion do not coincide with areas of enhanced optical reflectivity, which may be attributed to permanent surface-related oxygen loss, whereas the bulk can partially re-establish its oxygenation level after multiple EM runs. These observations highlight that optical microscopy by itself cannot adequately track bipolar electromigration, which is pertinent to device operation.

\begin{figure}[!ht]
    \centering
    \includegraphics[width=\linewidth]{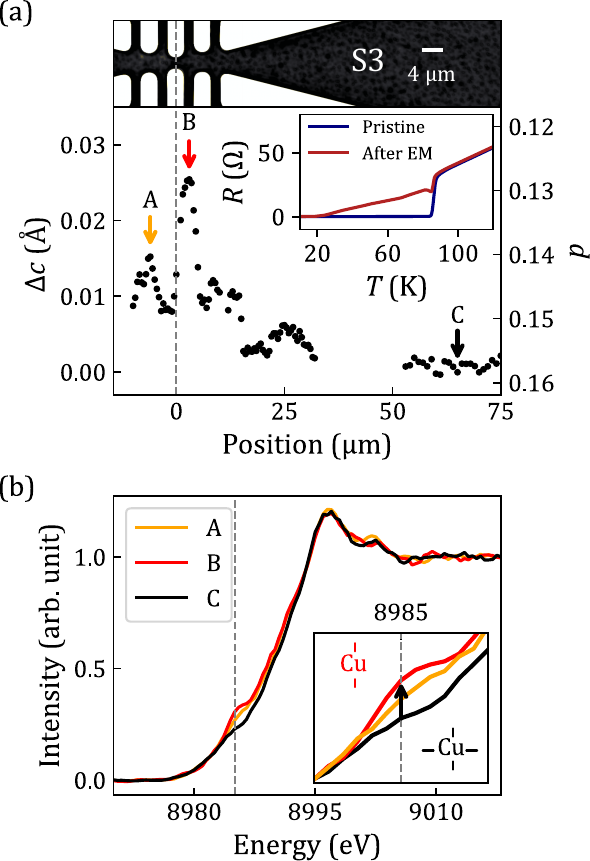}
   \caption{(a) Profile of the $c$-axis lattice-parameter variation measured by NanoXRD for sample S3. The upper panel shows a false-colored optical image of the device, aligned with the NanoXRD sampling positions. No increase in reflectivity is detected. The right y-axis indicates the corresponding hole doping $p$. The inset shows resistance versus temperature curves measured in the central constriction before and after EM.(b) Cu K-edge XANES spectra acquired from the regions marked by the colored arrows in (a), highlighting the correlation between structural modifications and copper coordination, manifested by changes in the pre-edge intensity.}
    \label{fig:4}
\end{figure}

With the aim of improving control over EM and the associated reflectivity changes, sample S3 was investigated at low temperature (150~K) in a vacuum cryostat, i.e., below room temperature but above its $T_c$, using a single EM run with a more limited resistance cutoff. Electromigration was halted when the central constriction reached a 4\% increase in resistance, reflecting the mild EM condition. NanoXRD data of S3, shown in Figure~\ref{fig:4}(a), reveals a pronounced maximum in the $c$-axis expansion at the right edge of the central bridge. The upper frame of the panel displays an optical image after EM, in which no contrast was detected. The limited range of $p$ observed in this sample allows us to estimate a threshold for the occurrence of significant optical changes, which appear when the local doping reaches $p \approx 0.12$. Additionally, the inset in Figure~\ref{fig:4}(a) presents resistance-versus-temperature curves before and after electromigration, showing a broadened superconducting transition in the central constriction with a locally reduced $T_c$, consistent with localized deoxygenation.

XANES spectra at selected regions of S3, in Figure~\ref{fig:4}(b), reaffirm a clear correspondence between atomic environment and local $c$-axis expansion, with point B corresponding to the maximum unit-cell expansion and exhibiting the most intense pre-edge feature. The lack of strong directional features, combined with low-temperature conditions, suggests that thermomigration contributed to vacancy redistribution alongside EM, likely driven by localized Joule heating and resulting temperature gradients. Another possible contribution is oxygen loss due to reduced oxygen partial pressure in the vacuum environment.

\begin{figure*}[!ht]
    \centering
    \includegraphics[width=1\linewidth]{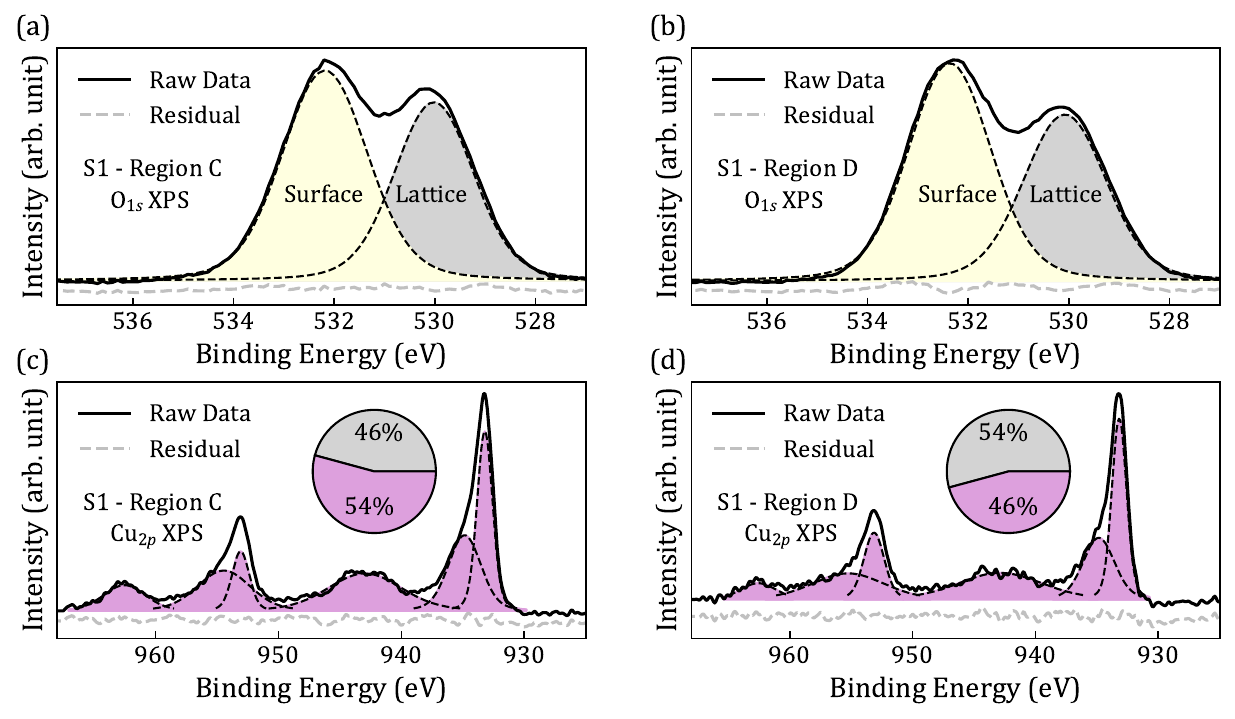}
    \caption{O 1$s$ and Cu 2$p$ XPS spectra for different regions of sample S1. Panels (a) and (b) show the O 1$s$ spectra, distinguishing surface and lattice oxygen, while panels (c) and (d) show the Cu 2$p$ spectra. Panels (a) and (c) correspond to region C, which exhibits increased optical reflectivity, and panels (b) and (d) correspond to region D, which shows no optical contrast. The pie charts indicate the relative fractions of lattice oxygen and copper in each region, revealing enhanced surface deoxygenation in the high-reflectivity area.}
    \label{fig:5}
\end{figure*}

In an attempt to directly assess the surface effects of EM, we performed XPS measurements on sample S1, focusing on regions C and D indicated by arrows in Figure~\ref{fig:2}(a). Again, as the sample was tilted, there is a geometric distortion along the axis of the longer dimension of the beam spot. However, for the selected points, outside the constriction regions, we ensured that the beam was fully on the sample surface, avoiding partial coverage or misalignment. We recall that region C corresponds to the area with increased reflectivity, whereas region D remains pristine. Figure~\ref{fig:5}(a–b) shows the O 1$s$ spectra for regions C and D, while Figure~\ref{fig:5}(c–d) presents the corresponding Cu 2$p$ spectra. While XPS is inherently surface-sensitive, probing only the top few nanometers of the film, the O 1$s$ spectra can be deconvoluted into contributions from lattice oxygen and adsorbed or surface-related species, enabling us to distinguish differences in local bonding environments rather than probing depth per se \cite{Wang2024}. The Cu 2$p$ core levels likewise serve as established markers of copper valence in perovskite oxides \cite{Nucker1995,Biesinger2017,Biesinger2011,Fenner1990}. Because the total area of each core-level feature is proportional to the amount of the corresponding atomic species within the probed volume \cite{shard2020}, the relative peak areas provide a semi-quantitative comparison between regions. The pie-chart insets in Figure~\ref{fig:5}(c-d) summarize the relative fractions of lattice oxygen and copper extracted from these fits, with the corresponding fractions ($f$) given by:

\begin{equation}
f_{\mathrm{O}} = \frac{\mathrm{O_{Lattice}}}{\mathrm{O_{Lattice}}+\mathrm{Cu}},
\end{equation}

\begin{equation}
f_{\mathrm{Cu}} = \frac{\mathrm{Cu}}{\mathrm{O_{Lattice}}+\mathrm{Cu}}.
\end{equation}

In the optical-contrast region C, the lattice-oxygen contribution drops to 46\%, compared with 54\% in the non-contrast region D, reflecting surface deoxygenation in the electromigrated area. For comparison, a similar analysis on corresponding regions of sample S3 yielded $f_{\rm O} = 49.5\%$ in the constriction-adjacent region, and $f_{\rm O} = 51.9\%$ in the farthest region, consistent with a largely pristine surface. These results further support the connection between optical contrast and surface oxygen depletion induced by electrical current.

\subsection{Conclusions}

Synchrotron-based X-ray diffraction and spectroscopy measurements indicate that, at spatial scales of hundreds of nanometers accessible to these techniques, oxygen in YBCO microbridge devices redistributes in a wave-like manner under high current densities, rather than along filamentary paths. Importantly, the deoxygenation signatures identified across different probes reinforce one another, most clearly in the spatial profile of unit-cell expansion, which mirrors the optical contrast observed after current-induced oxygen migration under single-polarity bias and ambient conditions. This correspondence confirms that both structural and optical measurements capture the same underlying oxygen redistribution, although optical microscopy alone is unable to reliably track bipolar electromigration effects---a condition relevant for device operation---as surface deoxygenation appears largely irreversible. Low-temperature operation with a reduced cutoff resistance provides a more controlled pathway, preventing local doping from reaching the estimated threshold $p \approx 0.12$, where significant optical modifications may occur. Under these conditions, thermally activated diffusion and possible oxygen loss due to the vacuum environment may have a larger impact. Altogether, these observations provide new insights into current-assisted oxygen migration mechanisms in YBCO thin film devices and underscore key aspects for employing this process as a reliable tool for tuning and controlling its properties.


\section*{Acknowledgements}

This research used facilities of the Brazilian Synchrotron Light Laboratory (LNLS) and the Brazilian Nanotechnology National Laboratory (LNNano), both part of the Brazilian Center for Research in Energy and Materials (CNPEM), a private non-profit organization under the supervision of the Brazilian Ministry for Science, Technology, and Innovations (MCTI). The CARNAÚBA and IPÊ beamlines staff (20241598 and 20241589), as well as the LCIS and MNF staff (20232334 and 20233770), are acknowledged for their assistance during the experiments. 

C.C.Q.G., E.A.A., S.M., D.A.D.C., P.S. and M.M. acknowledge financial support from the São Paulo Research Foundation (FAPESP, Grants 2023/11915-1, 2024/05377-0, and 2022/03124-1), the National Council for Scientific and Technological Development (CNPq, Grant 310514/2025-8), and Coordenação de Aperfeiçoamento de Pessoal de Nível Superior (CAPES) – Finance Code 001. P.S. acknowledges financial support from FAPESP through the Research, Innovation and Dissemination Center for Molecular Engineering for Advanced Materials – CEMol (Grant CEPID No. 2024/00989-7). This work was also supported by the INCT project Advanced Quantum Materials, involving the Brazilian agencies CNPq (Proc. 408766/2024-7, and 302786/2025-2), FAPESP (Proc. 2025/27091-3), and CAPES. A.V.S. and A.P. acknowledge COST (European Cooperation in Science and Technology) through COST Action SUPERQUMAP (CA21144). A.P. also acknowledges financial support from MICIU/AEI /10.13039/501100011033/ through CEX2023-001263-S, PID2021-124680OB-I00, PID2024-156025OB-I00, co-financed by ERDF A way of making Europe, and from the Spanish Nanolito network (RED2022-134096-T).

The authors thank Daniel Stoffels for valuable discussions.

\section*{Supporting Information}

The Supporting Information provides additional experimental details, including the procedures for thin film deposition and the basic characterization of the samples. It also presents the peak-fitting methodology applied to the NanoXRD data and the analysis of beam-induced damage effects.

\printbibliography
\newpage

\end{document}